# Suppressing Chemical Corrosions of Lithium Metal Anodes

Bingyu Lu[1], Weikang Li[1], Diyi Cheng[2], Miguel Ceja[1], Wurigumula Bao[1*], Chengcheng Fang[3*], Ying Shirley Meng[1,4*]

[1]Department of NanoEngineering, University of California, San Diego, La Jolla, CA 92093, USA
[2]Materials Science and Engineering Program, University of California, San Diego, La Jolla, CA 92093, USA
[3]Department of Chemical Engineering and Materials Science, Michigan State University, East Lansing, MI 48824, USA.
[4]Pritzker School of Molecular Engineering, University of Chicago, Chicago, IL 60637

*Correspondence to: shirleymeng@uchicago.edu, cfang@msu.edu, wubao@eng.ucsd.edu.

**Abstract**: The lithium (Li) metal anode is essential for next generation high energy density rechargeable Li metal batteries. Although extensive studies have been performed to prolong the cycle life of Li metal batteries, the calendar life, which associates with chemical corrosion of Li metal in liquid electrolytes, has not been quantitatively understood. Here, by combing the Titration Gas Chromatography (TGC) method and Cryogenic Focused Ion Beam (Cryo-FIB), we established a quantitative relationship between the chemical corrosion rate and electrochemically deposited Li morphology in various liquid electrolyte systems. We identified that the corrosion rate is dominated by the porosity of the deposited Li. The larger the porosity of deposited Li has, the faster the corrosion rate will be. We further proposed strategies to mitigate the chemical corrosion on Li thus to extend the calendar life of Li metal batteries. By strictly controlling the stacking pressure during Li plating, Li deposits with ultra-low porosity can be achieved, suppressing the corrosion rate to 0.08 ± 0.16%/day compared with 1.71 ± 0.19%/day of the high-porosity Li.

**Introduction**

With the growing demand of high performance electric vehicles and personal portable devices, lithium (Li) metal anode is crucial for developing high energy density rechargeable batteries (> 500 Wh/kg) due to its high specific capacity (3,860 mAh/g) and low electrochemical potential (–3.04 V versus the standard hydrogen electrode)[1–3]. Although extensive studies have been performed to overcome dendrite growth and low coulombic efficiency problems associated with the practical use of Li metal anode[4–6], there is a lack of comprehensive understanding of the stability and storage properties of lithium metal anodes in liquid electrolytes, where corrosion plays a critical role[7,8]. Corrosion is a common chemical/electrochemical process that almost all metals will experience when exposed to an oxidating environment[9,10]. During the corrosion, the fresh metal surface will be oxidized, followed by the formation of its corresponding ionic species and the release of electrons[11,12]. The corrosion is usually accompanied by the generation of a passivating layer which blocks the transfer of the electrons and eventually stops the continuous oxidization of the metal[13]. Without the formation of this passivating layer, the corrosion of the metal will continue until the thermodynamic equilibrium is reached when electrochemical potential (μ) difference between the metal and the environment becomes zero[14,15]. In an electrochemical cell, the electrode materials are usually immersed in ion-conducting solvents, which will allow corrosions to take place[16]. Fortunately, in the traditional Li ion batteries (LIBs), a dense and passivating solid–electrolyte interphase (SEI) can be quickly formed during the first a few cycles[17,18], thus preventing the following corrosion of the electrode materials, which enables a stable cycling and storage life for the LIBs.

Several works have shown that the galvanic (electrochemical) corrosion of Li metal will happen when the interface between Li metal and current collector is exposed to the liquid electrolyte[19–21], where the exchange of charges will take place[22]. Kolesnikov A. and his co-workers monitored the morphological change of the Li powders casted on the Cu substrate and discovered a continuous shrinkage of the Li near the Li||Cu interface, which was believed to be caused by the Galvanic corrosion between Li and Cu. However, this Li metal and current collector interface can be largely blocked from the electrolyte if a dense Li morphology is achieved during Li plating[5]. Therefore, more research work should be focusing on the chemical stability between Li metal and the liquid electrolytes. It is well known that because of the extremely low standard redox potential of Li (–3.04 V versus the standard hydrogen electrode), Li can immediately react with essentially any electrolyte upon contact[23,24]:

$$n\,Li \rightarrow n\,Li^+ + n\,e^-$$

$$Electrolyte + n\,e^- \rightarrow SEI + soluble\ electrolyte\ decomposition\ products$$

This simplified reaction pathway depicts the spontaneous chemical corrosion process between Li and the liquid electrolyte. The SEI layers formed during this process, which usually consist of $Li_2O$, LiF, and other organic compounds, can quickly passivate the Li metal surface[25]. However, due to the inhomogeneity in the solubility and electronic conductivity of these SEI components[26], the continuous decomposition of the electrolyte might still occur on the SEI surface, which is observed in the recent Cryo-STEM work done by Boyle D. T. and his colleagues[27], leading to the further corrosion of the Li metal even after the formation of the initial passivating (SEI) layer. As a result of this continuous chemical corrosion, the corroded Li metal anode will suffer from a loss of active $Li^0$ material, an increase of cell impedance, and eventually a poor cycling/storage life, which will cause the failure of the cell.

The lifetime of a battery system depends on two key factors: 1) calendar life, the degradation over storage time and 2) cycle life, the degradation over charge-discharge cycles[28]. Despite the fact that tremendous amount of work has been done trying to extend the cycle life of Li metal batteries, few previous works have explicitly considered the key parameters in determining the calendar life of Li metal batteries, as well as the tradeoffs between the calendar life and cycle life. In the extreme case of primary Li battery, where calendar life is the more preferential factor than cycle life, many work has been done to protect the Li metal from degrading during storage by constructing a stable SEI layer for Li protection[29]. However, such artificial SEI can be easily destructed during cycling because of the large volume change of Li plating/stripping. Therefore, it is crucial that the reactivity of the Li is strictly controlled when designing a secondary battery system so that lifetime of it is not limited by neither the calendar life nor the cycle life. Thus, this work focuses on deciphering the key parameters in determining the rate of the chemical corrosion (chemical reactivity) of Li metal in the liquid electrolyte during the extended resting period and find ways to prolong the calendar life of the rechargeable Li metal batteries.

Here, by using the Titration Gas Chromatography (TGC) method[30], we quantify the corrosion trend of plated Li in four representative liquid electrolyte systems: high concentration ether based "Bisalt" electrolyte (4.7m LiFSI + 2.3m LiTFSI in DME)[31], low concentration ether based "Nitrate" electrolyte (1M LiTFSI in DME:DOL with 2wt% LiNO$_3$), carbonate based "Gen 2" electrolyte (1.2M LiPF$_6$ in EC:EMC) and Localized High Concentration Electrolyte (LHCE, LiFSI:DME:TTE in molar ratio of 1:1.2:3)[32]. These four electrolytes are used in this work because they represent the four most popular electrolyte systems studied in the field[33]. The morphological changes of the Li at different stages of corrosion are also recorded by Cryo-FIB/SEM[34]. It is found that the porosity of the plated Li has a significant effect on determining the corrosion rate of the Li in liquid electrolyte (**Fig 1**). By combining the TGC method and also the three-dimensional (3D) reconstruction of the plated Li by Cryo-FIB/SEM, the porosity of the plated Li is quantified and its corresponding corrosion rate in the liquid electrolyte is calculated. Finally, by using the advanced LHCE and optimized stacking pressure (350 kPa)[35], a ultra-low porosity of Li is achieved. The resulting low porosity Li experiences only 0.8% loss of Li$^0$ mass after 10 days of immersion in the liquid electrolyte. The fundamental correlation among Li metal porosity, SEI composition and the Li corrosion rate is revealed in this work.

**Quantifying Li metal corrosion rates in liquid electrolytes by TGC**

First, 0.318 mAh of Li is plated onto a 1.27 cm$^2$ of Copper (Cu) substrate at a rate of 0.5 mA/cm$^2$ in a coin cell setup with 55 µL of Bisalt, Nitrate and Gen 2 electrolyte. After plating, the deposited Li metal with the Cu substrate is kept in coin cell with the corresponding electrolytes with the open circuit condition (without linked to a cycler). After a certain period of storage time, the cells are disassembled and TGC method is used to quantify the mass of metallic Li$^0$ remained on the Cu substrate (**Fig. 1a**). The Li mass retention (%) as a function of storage time is shown in **Fig. 1b**, 1f and 1j. For Li plated in all three electrolytes, there is a sudden drop of Li$^0$ mass in the first 24 hours of resting (**Fig. 1c, 1g and 1k**). A similar trend is also found by Boyle D. T. and his colleagues in their recent work[27], where a fast drop of second cycle coulombic efficiency (CE) is observed after 24 hours of cell resting. After that, the corrosion rates of Li plated in Bisalt and Nitrate electrolytes relatively slow down, which fluctuate at around 0.5 µg/day during 5 weeks of resting. However, a more drastic loss of Li$^0$ mass is observed in the Gen 2 electrolyte. The continuous corrosion of Li causes a 60.8% loss of Li$^0$ mass after 5 weeks of resting, especially after 2 weeks of resting when a sudden increase of corrosion rate is observed (**Fig. 1j**).

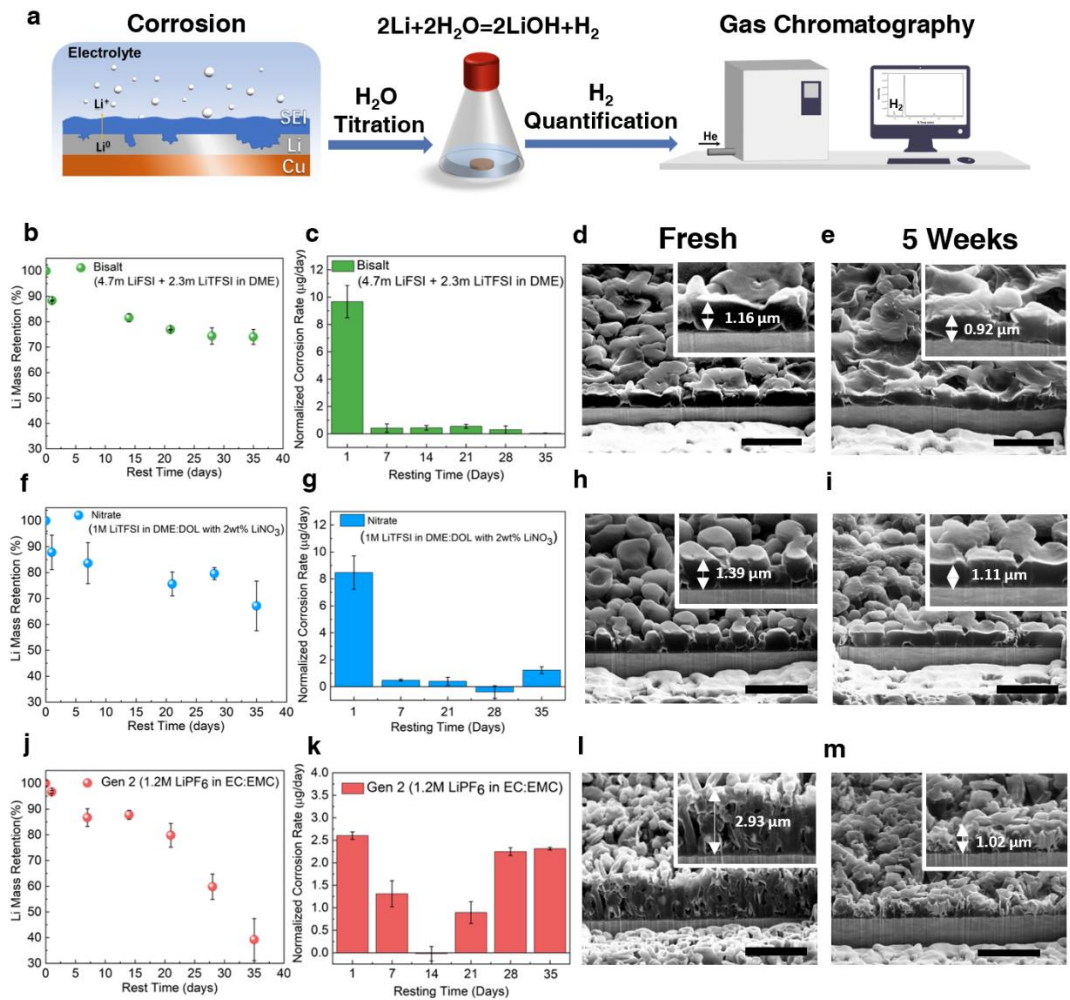

**Figure 1. The trends of Li metal corrosion in liquid electrolytes.** (a) the quantification of Li corrosion by the TGC method. (b-e) The trend of Li metal corrosion in high concentration ether based "Bisalt" electrolyte: (b) the Li mass retention (%) and (c) the normalized corrosion rate of Li (µg/day) as a function of resting time; The SEM images of the Li morphology (d) before and (e) after 5 weeks of resting. (f-i) The trend of Li metal corrosion in ether solvent based "Nitrate" electrolyte: (f) the Li mass retention (%) and (g) the normalized corrosion rate of Li (µg/day) as a function of resting time; The SEM images of the Li morphology (h) before and (i) after 5 weeks of resting. (j-m) The trend of Li metal corrosion in carbonate solvent based "Gen 2" electrolyte: (j) the Li mass retention (%) and (k) the normalized corrosion rate of Li (µg/day) as a function of resting time; The SEM images of the Li morphology (l) before and (m) after 5 weeks of resting. Total amount of 0.318 mAh of Li is plated at a rate of 0.5 mAh/cm$^2$ with 55 µL of electrolyte amount in all samples. The scale bars represent 5 µm.

The morphological study by Cryo-FIB/SEM further confirmed the corrosion trend obtained by TGC. The Li morphology in Bisalt and Nitrate electrolyte does not experience a significant change before and after 5 weeks of resting in open circuit (**Fig 1d-e and 1h-i**). The Li retains mostly its granular shape even after the resting period, although there is a decrease of Li thickness in both cases, which corresponds to the loss of Li$^0$ mass. As expected, a sharp change in the morphology of Li plated in the Gen 2 electrolyte is observed by Cryo-FIB/SEM (**Fig. 1l-m**). The freshly deposited Li in Gen 2 electrolyte exhibits a whisker-like morphology, while the corroded Li shows a porous and powder-like morphology. There is also a substantial decrease of Li thickness from 2.93 µm to 1.02 µm, which again agrees with the loss of $Li^0$ mass quantified by TGC. The three different electrolyte systems, Bisalt, Nitrate and Gen 2, show two drastically different corrosion trends: Bisalt and Nitrate electrolytes show a fast corrosion rate during the first 24 hours, but stabilizes quickly afterward, while Gen 2 shows a continuous corrosion rate throughout the 5 weeks of resting period. A similar corrosion trend study is also done for commercial ultra-thin (50 µm) Li foils (**Fig. S1**). It is found that the 50 µm Li foil without Cu substrate lost more active $Li^0$ mass after one week of immersion than that of the one with Cu substrate, because it has more contact surface area with the electrolyte, further confirming the hypothesis that once the Li||Cu interface is blocked from the electrolyte, the galvanic corrosion will not take place in an extensive rate. Whereas in the case of electrochemically deposited Li, since the electrolyte systems tested so far are based on fundamentally different chemistries, the resulted Li might have different SEI components and porosity. Therefore, there might be two possible reasons for the difference in the corrosion trends from the three electrolyte systems: 1) Li metal surface chemistry (SEI); 2) Li metal porosity.

To elucidate either surface chemistry or porosity dominates the corrosion rate resulting from the three electrolyte systems, X-ray Photoelectron Spectroscopy (XPS) with depth profiling is performed on the freshly deposited and corroded Li metal to verify the interphases chemical information (**Fig. 2**). We identified that the fresh SEI components in all three electrolyte systems are almost identical. In both Bisalt and Nitrate samples, the change of C-F implies the presence of Li salt (LiTFSI and LiFSI) on the surface of the Li, which can be confirmed by 533 eV peak in O 1s and (C-F at 293 eV and C-S=O at 289 eV) eV peaks in C 1s as well. Other than the Li salts, Li-F, $Li_2O$ and typical organic species such as C-O/C=O/ROLi can be well located, which is consistent with previous literature results[30,32] However, the SEI layer of the Nitrate samples seems to be thinner than that of Bisalt, because there is $Li^0$ signal can be detected in the Nitrate samples even before etching. **Fig. 2b** demonstrates the results from Gen 2 sample, which is quite like the Bisalt samples considering the major components include Li-F, $Li_2O$ and typical organic species in the freshly deposited samples. After 3 weeks of resting, in the Bisalt samples, there is an accumulation of Li salt in the surface of the Li, while other SEI components still preserve well in the surface layers. Similar to the Bisalt case, SEI components in the Nitrate sample also preserve well after 3 weeks of resting. However, in the Gen 2 case, the $Li_2O$ content mostly disappears after 3 weeks of resting. The surface layers mainly consist of LiF after the corrosion. The results of Gen 2 samples show that the SEI layers have undergone a significant change during the resting period (highlighted by blue and violet hades), which correspond to the corrosion trend of the Li deposited in Gen 2 where a fast corrosion rate is observed throughout the 5 weeks of resting (**Fig. 1e**). Based on the results so far, it can be observed that in the case of Bisalt and Nitrate electrolytes, where ether-based solvents are used, the SEI components are much more stable than that of the Gen 2 electrolyte. The stability of the SEI also reflected in the slow corrosion trend in the Bisalt and Nitrate samples. Therefore, we believe that the SEI layers need to be strong enough to survive the corrosion process to mitigate the continuous mass loss of Li in liquid electrolyte.

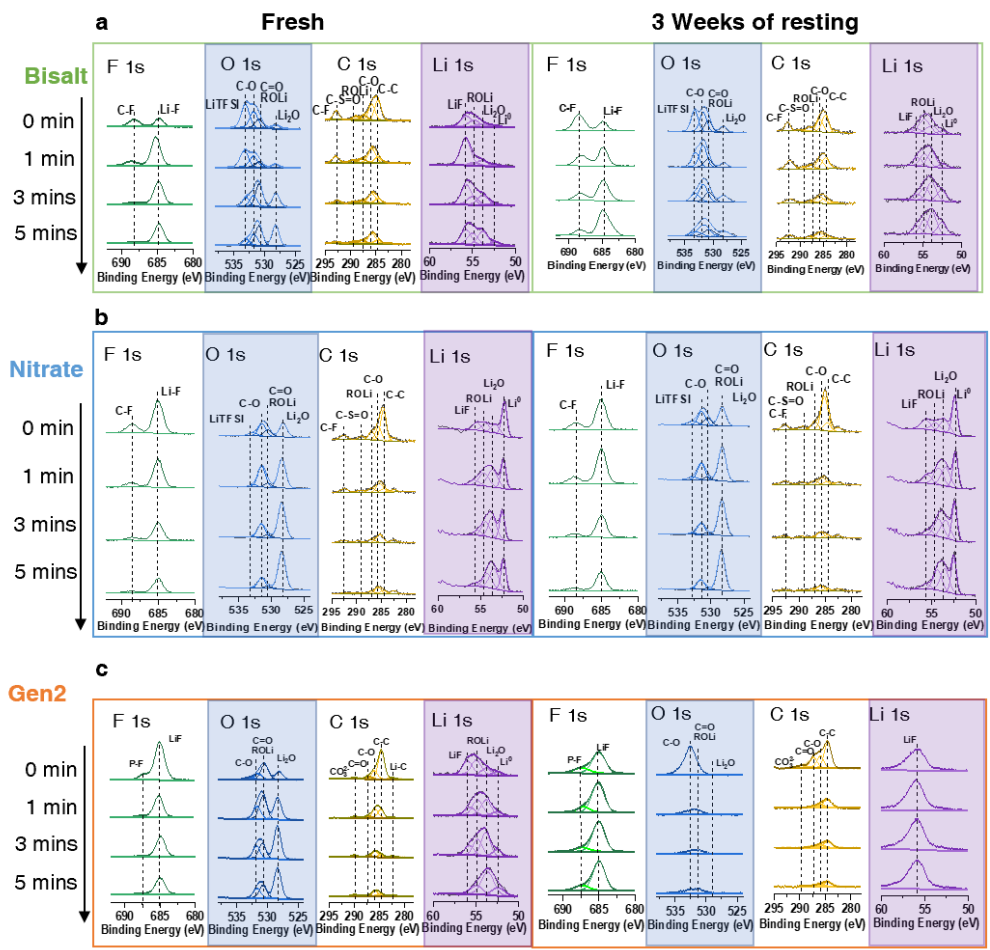

**Figure 2. The XPS depth profiling of deposited Li metal.** Chemical evolution of F 1s, O 1s, C 1s, and Li 1s of (a) Bisalt electrolyte and (b) Nitrate electrolyte (c) Gen 2 electrolyte before and after 3 weeks of resting in its corresponding electrolyte.

**Effect of porosity on the Li corrosion rate**

Our previous work has shown stacking pressure could effectively control the porosity of deposited Li[35]. To investigate the effects of Li porosity on the corrosion rate, a split cell together with a pressure sensor is used for controlling the stacking pressure during Li plating (**Fig. 3a**), which can help us to obtain deposited Li with different porosities. Similar to the previous case, 0.318 mAh of Li is plated onto Cu substrate at a rate of 0.5 mAh/cm$^2$ in Gen 2 electrolyte. However, the electrode size is changed from 1.27 cm$^2$ to 0.385 cm$^2$ to accommodate the smaller size of the split cell. Two different stacking pressures, coin cell pressure (150 kPa) and optimized Li plating pressure (350 kPa), are applied during the Li plating process. The plated Li with the Cu substrate is recovered from the split cell and immersed in flooded electrolyte (~1mL) for corrosion study. **Fig. 3b** shows the Li$^0$ mass retention (%) of the Li plated under two pressures. The lower porosity of the Li plated under 350 kPa helps to suppress the Li corrosion rate. The metallic Li$^0$ lost about 18% of its original mass as a contrast to 29.2% in the case of coin cell pressure. When comparing the improved Li corrosion trend to that of Li plated in Bisalt electrolyte under coin cell pressure, it is found that the two corrosion trends are similar (**Fig 3c**), meaning that with the optimized stacking pressure applied during Li plating, the resulting low porosity Li can have a limited corrosion rate even in conventional carbonate electrolyte. The cryo-FIB/SEM images also illustrated the morphological change of the Li plated under different pressures before and after corrosion. From the top-view and cross-sectional images of the Li plated under coin cell pressure, the resulting Li is in whisker-like morphology (**Fig. 3d, 3h**). However, just after 7 days of immersion in Gen 2 electrolyte, there is a noticeable shrinkage in both the plated Li's thickness and the diameter of whiskers. On the contrary, the morphology of Li plated under 350 kPa pressure did not show a significant change. The Li retained its dense morphology after 7 days of immersion, especially in the cross-section. The only noticeable change happens on the top surface of the dense Li where some flower-like materials begin to form on the surface. The effect of porosity on corrosion rate for low concentration ether-based electrolyte (LCE,1M LiTFSI in DME:DOL) is also conducted (**Fig. S2**). A similar fast corrosion trend of the Li plated in the Gen 2 (**Fig S2a, S2d**) is also observed in the Li plated in the LCE, where porous Li whiskers are grown (**Fig S2b, S2e**). With the results so far, it can be seen that the Li corrosion only takes place at the interface between Li and the electrolyte. Even when Li is deposited in the ether-based electrolyte (LCE), which is believed to generate more stable SEI, because of the high porosity of the deposited Li, the corrosion rate of Li in LCE is considerably higher than that of the Bisalt and Nitrate electrolytes. Therefore, the porosity of the plated Li should play a major role in controlling the corrosion rate of Li.

To validate if the Li electrode porosity is the dominating factor of corrosion rate, we selectively deposited Li in Gen 2 electrolyte, with a range of stacking pressure applied during Li plating to plate Li with different porosities. Cryo-FIB/SEM is used to obtain the 3D reconstruction of the plated Li and calculate its porosity (**Fig. 3l-o**). The as-plated Li is then immersed into the Gen 2 electrolyte and rest for three weeks. The $Li^0$ mass loss is used for calculating the corrosion rate of the Li using the following equation[15]:

$$R = \frac{\Delta m}{A_S T}$$

where $\Delta m$ is the change in mass, $A_s$ is the electrode area, and T is the corrosion time. The detailed calculated parameters are listed in Table S1 in the supporting information. As shown in **Fig. 3p**, the calculated Li corrosion rate directly correlates with the plated Li's porosity. The results further validated our hypothesis that the porosity of the plated Li is the major factor in determining the Li corrosion rate.

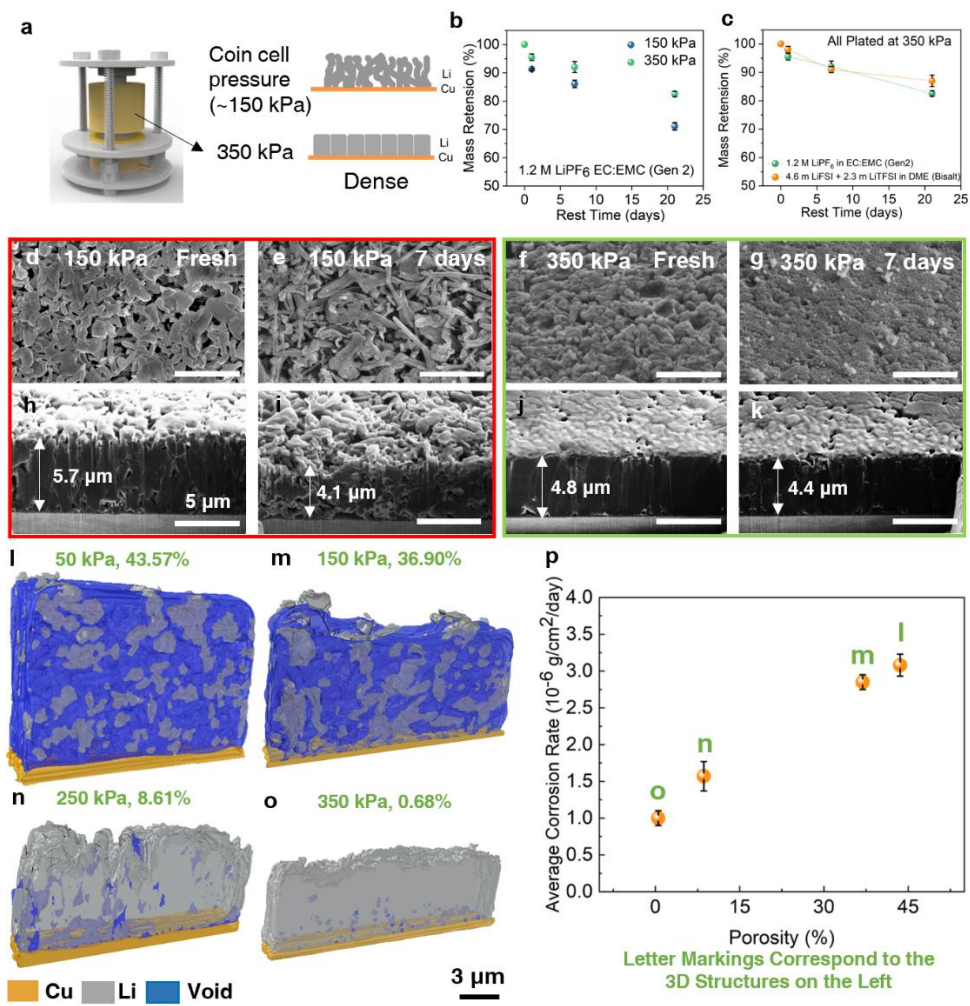

**Figure 3. The effect of morphology control on limiting the corrosion of Li in liquid electrolyte.** a) Schematics of stacking pressure control set up. (b) Trend of Li$^0$ mass retention (%) of Li plated under 2 different stacking pressure (coin cell pressure and 350 kPa pressure) in Gen 2 electrolyte. (c) The comparison between Li$^0$ mass retentions (%) of Li plated in Gen 2 electrolyte with 350 kPa and in Bisalt electrolyte in coin cell. (d-k) The top-view and cross-sectional SEM images of the deposited Li metal in Gen 2 electrolyte: under coin cell pressure after (d, h) freshly deposited, (e, i) after 7 days of resting and under 350 kPa pressure after (f, j) freshly deposited, (g, k) after 7 days of resting. The 3D reconstruction of deposited Li metal plated under different stacking pressures resulting in different Li porosities: (l) 50 kPa and 43.57% porosity; (m) 150 kPa and 36.90% porosity; (n) 250 kPa and 8.61% porosity and (o) 350 kPa and 0.51% porosity. The Li metal corrosion rate as a function of Li porosity: (p) The Li metal corrosion rate and its correlation with the porosity of the freshly deposited Li. All Li is plated in Gen 2 electrolyte. Total amount of 0.318 mAh of Li is plated at a rate of 0.5 mA/cm$^2$ for all samples.

**Suppressing Li metal corrosion: advanced electrolyte and optimal stack pressure**

With the results so far, it can be concluded that the controlling interface between Li and electrolyte is the key in suppressing the chemical corrosion of Li in liquid electrolytes: the contact area needs to be minimized while maintaining a robust SEI layer. To further demonstrate the feasibility of limiting the Li corrosion by controlling the porosity of Li, a novel electrolyte system, the localized high concentration electrolyte (LHCE)[32], with LiFSI:DME:TTE in a molar ratio of

1:1.2:3 is used to plate Li under different stacking pressures. By applying the optimized stacking pressure of 350 kPa during Li plating, ultra-low porosity was achieved for deposited Li, which results in less than 0.8% of metallic Li loss after 10 days of resting in LHCE electrolyte (**Fig. 4a**). The corrosion of the metallic Li is significantly mitigated by minimizing the contact area between Li and electrolyte. As shown in **Fig. 4b** and 4c, the dense Li morphology is well retained even after 21 days of immersion in the LHCE electrolyte. In addition to the dense morphology, the SEI layers of the Li deposited in LHCE is also quite stable. After three weeks of resting, there is no observable changes in the SEI components (**Fig. S5**). Even at elevated temperature of 40°C, the corrosion rate of Li is lower than that of Gen 2 electrolyte (**Fig. S4**). However, when the resting ambient temperature is raised to 55°C, a fast corrosion rate is observed for LHCE (**Fig. S4**). More work needs to be conducted to find out ways to mitigate the corrosion of Li metal at elevated temperatures.

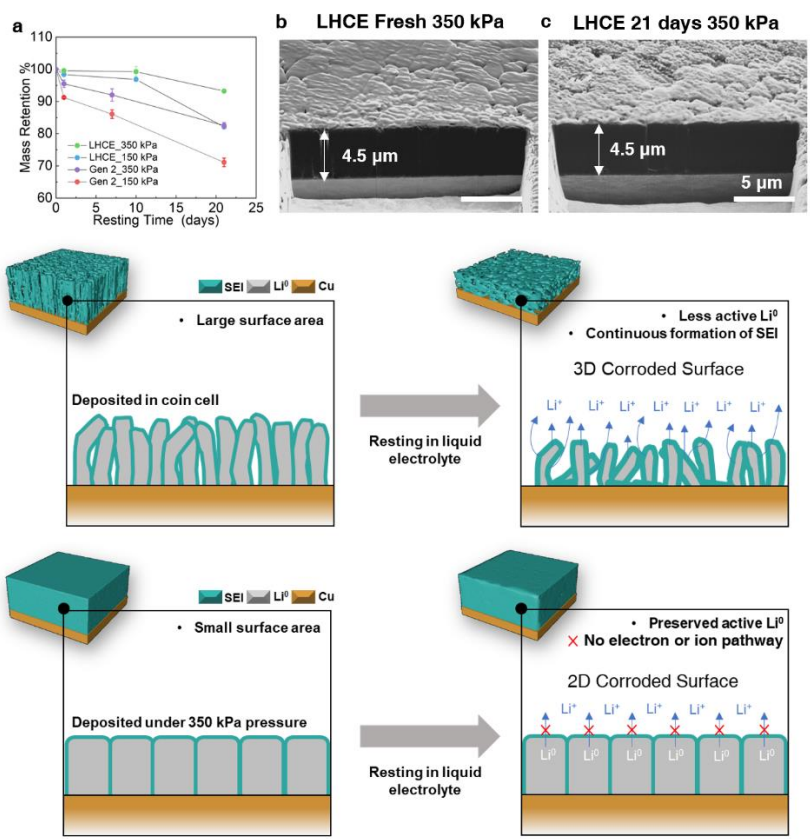

**Figure 4. Controlling the corrosion of Li metal in liquid electrolytes**. (a) Trend of $Li^0$ mass retention (%) of Li deposited in Gen 2 and LHCE (LiFSI:DME:TTE in molar ratio of 1:1.2:3) under coin cell and 350 kPa pressure. Cross-sectional images of Li plated in LHCE under 350 kPa pressure (b) after freshly deposited and (c) after 21 days of resting. (d) Schematics of possible ways to suppress the Li metal corrosion in liquid electrolyte. Total amount of 0.318 mAh of Li is plated at a rate of 0.5 mA/cm$^2$ in all samples. The schematics of the relationship between the Li morphology and its corrosion trend. The high porosity Li whiskers plated in coin cell setup will have a high contact surface area with the liquid electrolyte, inducing continuous chemical corrosion of Li. The low porosity Li plated under optimized stacking pressure has only a 2D contact surface area with the liquid electrolyte, which largely limits the chemical corrosion of Li in the liquid electrolyte.

Overall, the strategies of suppressing the Li corrosion in liquid electrolytes should focus on three parts. First and most importantly, an optimized stacking pressure should be applied during Li plating to achieve a dense and uniform Li morphology. Since the corrosion of Li requires contact with the liquid electrolyte, by limiting the surface area (porosity) of the plated Li, the corrosion rate of Li during resting period can be effectively mitigated. Second, a dense and stable interface should be constructed during plating, which can be achieved mainly by using an advanced electrolyte system such as LHCE. A stable interface can further block the charge exchange between Li and the liquid electrolyte, thus limiting the corrosion of the plated Li. Lastly, more work should be done to design a dense but flexible surface coating that can be tightly applied on Li surface and retain its structural integrity during cycling[36], and possibly further mitigate the corrosion of Li metal at elevated temperatures. In this way, the total blockage of charge transfer between Li and liquid electrolyte can be achieved throughout the resting period, and the corrosion of Li metal can be further mitigated.

**Conclusion**

We systematically studied the corrosion rate of Li in liquid electrolytes regarding: 1) Li surface chemistry; and 2) Plated Li porosity. For the Li surface chemistry, it was found that the major SEI components mainly consist of LiF, $Li_2O$, and organic Li containing species for plated Li in both ether and carbonated based electrolytes. The plated Li with a well-controlled porosity shows drastically decreased corrosion rate even in Gen 2 electrolyte, illustrating that the most crucial parameter in determining the corrosion rate of Li is the contact surface area between Li and liquid electrolyte, which is the porosity of the Li. By using advanced electrolyte (LHCE) and optimized stacking pressure (350 kPa) for Li deposition, the ultra-low porosity Li is plated, which only has a 2D contact surface area with the liquid electrolyte. The low contact surface area helped the dense Li to stabilize in the liquid electrolyte and the Li loses only about 0.8% of its active mass after 10 days of immersion in liquid electrolyte. The work here has shown that by controlling the contact area between Li and liquid electrolyte can help to effectively suppress the corrosion of Li.

**Acknowledgments:**

**Funding:** This work was supported by the Office of Vehicle Technologies of the U.S. Department of Energy through the Advanced Battery Materials Research (BMR) Program (Battery500

Consortium) under Contract DE-EE0007764. Cryo-FIB was performed at the San Diego Nanotechnology Infrastructure (SDNI), a member of the National Nanotechnology Coordinated Infrastructure, which is supported by the National Science Foundation (grant ECCS-1542148). We acknowledge the UC Irvine Materials Research Institute (IMRI) for the use of the XPS, funded in part by the National Science Foundation Major Research Instrumentation Program under Grant CHE-1338173. The authors would like to acknowledge Neware Technology Limited for the donation of BTS4000 cyclers which are used for testing the cells in this work.

**Author Contribution**

B.L., W.B., C.F. and Y.S.M. conceived the ideas and designed the experiments. B. L. and M.C. implemented the electrochemical tests. B.L. performed the cryo-FIB experiments. B.L., W.B., and M.C. performed the TGC analysis. W.L. performed XPS analysis. B.L. and W.B. wrote the manuscript. All the authors discussed the results and commented on the manuscript. All the authors gave approval to the final version of the manuscript.

**Competing interests**

The authors declare no competing interests.

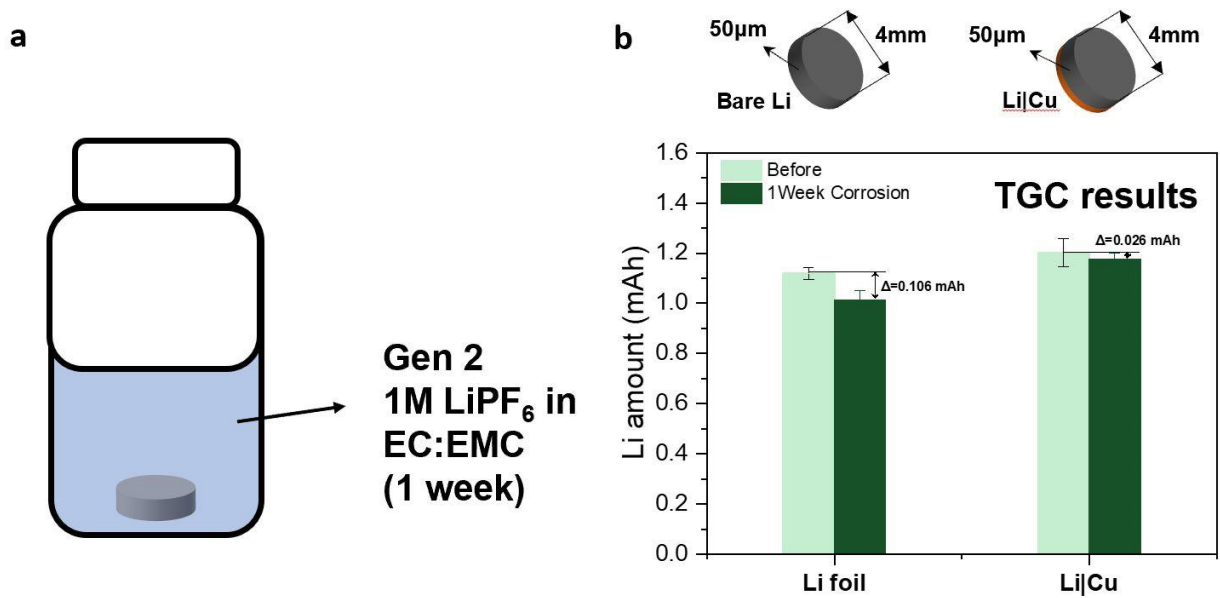

**Figure S1. The effect of Cu substrate.** (a) schematics of the corrosion test of the pristine Li foil; (b) the quantification results of Li foil before and after immersion in Gen 2 electrolyte (1M $LiPF_6$ in EC:EMC) after one week. Two types of 50 µm thin Li foil, with and without Cu substrate, were punched into 4mm pieces. The punched thin Li foils were then immersed in Gen 2 for one week before taking TGC measurement. It was shown that the Li foil without Cu substrate lost more active $Li^0$ mass after one week of resting, because it has more contact surface area with the electrolyte.

**Table 1.** The detailed quantification of $Li^0$ mass loss and its corresponding corrosion rate.

| Sample | Porosity | Mass loss (After 3 weeks) | Corrosion rate (g/cm$^2$/day) |
|---|---|---|---|
| High porosity (Coin cell) | 43.57% | 2.49*10$^{-5}$ g | 3.08*10$^{-6}$ |
| Medium porosity (150 KPa) | 36.90% | 2.30*10$^{-5}$ g | 2.85*10$^{-6}$ |
| Low porosity (250 KPa) | 8.61% | 1.27*10$^{-5}$ g | 1.57*10$^{-6}$ |
| Dense (350 KPa) | 0.51% | 0.81*10$^{-5}$ g | 1.00*10$^{-6}$ |

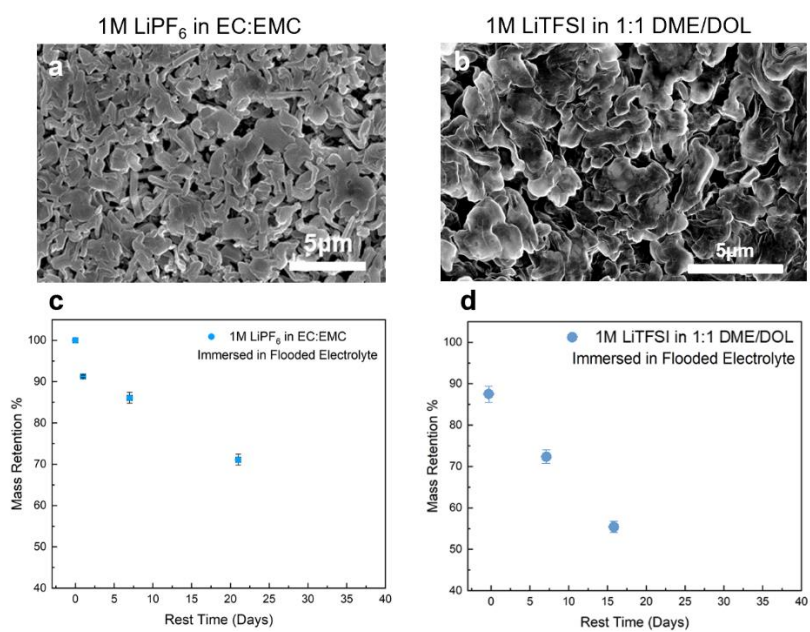

**Figure S2**. **The effect of surface area on corrosion rate.** The top-view SEM images of plated Li in (a) Gen 2 electrolyte; (b) 1M LiTFSI in DME:DOL; The $Li^0$ mass retention (%) as a function of resting time in (c) Gen 2 electrolyte; (d) 1M LiTFSI in DME:DOL; Total amount of 0.318mAh of Li is plated at a rate of 0.5mAh/cm$^2$ in all samples.

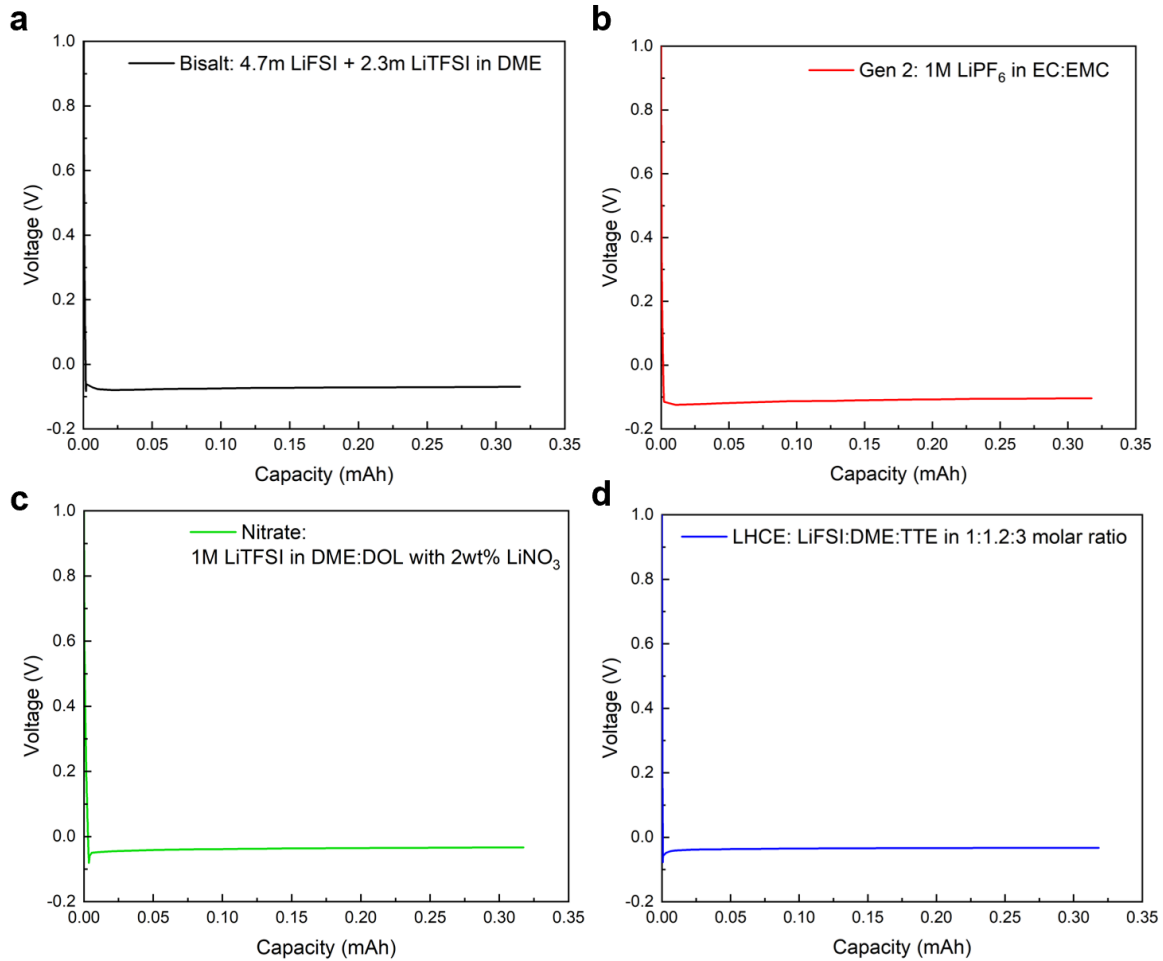

**Figure S3. The voltage profiles of Li plating.** (a) in Bisalt electrolyte; (b) in Gen 2 electrolyte; (c) in Nitrate electrolyte; (d) in LHCE. Total amount of 0.318mAh of Li is plated at a rate of 0.5mAh/cm$^2$ in all samples.

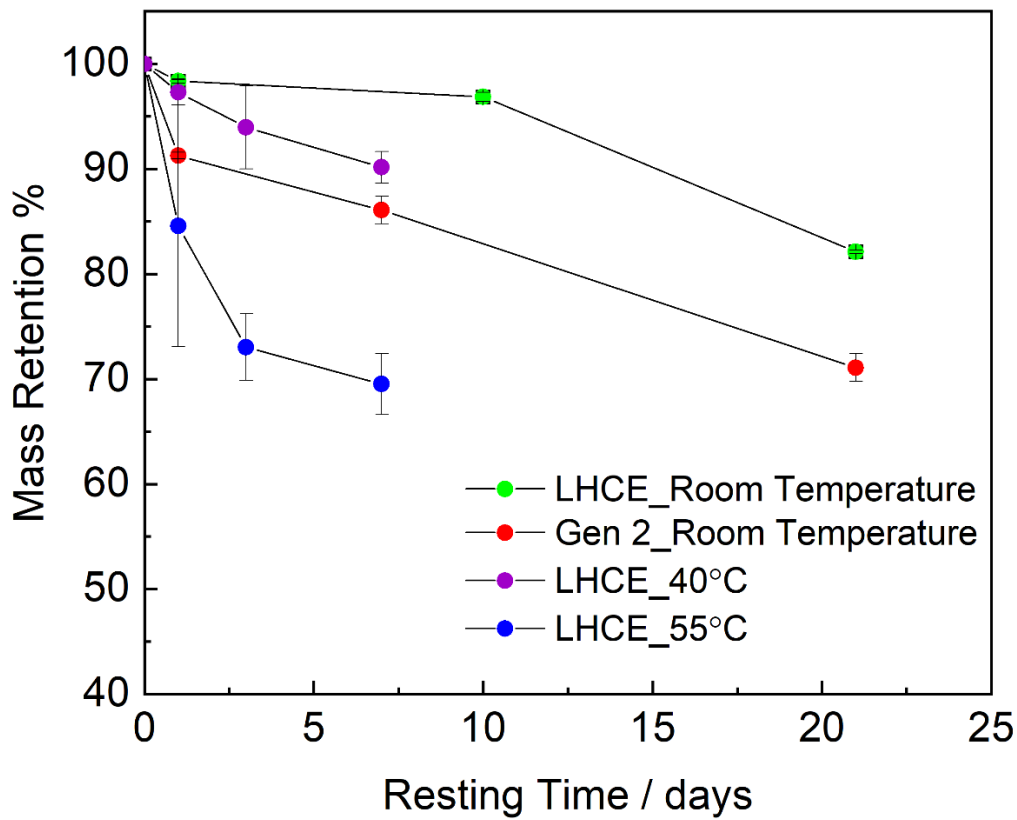

**Figure S4.** Trend of Li$^0$ mass retention (%) of Li deposited in Gen 2 and LHCE at various temperatures.

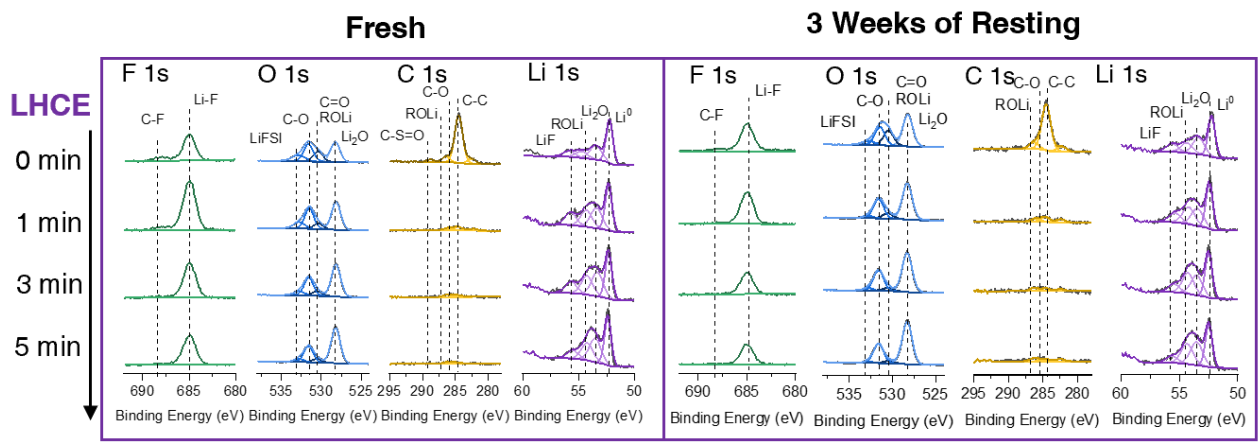

**Figure S5.** The XPS depth profiling of F 1s, O 1s, C 1s, and Li 1s of Li deposited in LHCE before and after 3 weeks of resting.